\documentclass[aps,prx,twocolumn,superscriptaddress,longbibliography]{revtex4-2}

\usepackage{epsfig}
\usepackage{epstopdf}
\usepackage{float}
\usepackage[caption=false]{subfig}
\usepackage[section]{placeins}
\usepackage{graphicx}		
\usepackage{epstopdf}
\usepackage{url}
\usepackage{amsmath}
\usepackage{mathtools}
\usepackage{enumitem}
\usepackage{amssymb}
\usepackage{physics}
\usepackage{breqn}
\usepackage{hyperref}

\usepackage{import}
\usepackage{tkz-euclide}
\usepackage{color}
\usepackage{soul}
\usepackage{dcolumn}
\usepackage{graphicx}
\usepackage{amsmath}
\usepackage{amssymb}
\usepackage{bm}
\usepackage{tikz}
\usepackage[section]{placeins}
\usetikzlibrary{decorations.pathmorphing,patterns}
\usepackage[capitalise]{cleveref}

\makeatletter
\let\cref@old@eq@setnumberOld\eq@setnumber
\def\eq@setnumber{%
	\cref@old@eq@setnumberOld%
	\cref@constructprefix{equation}{\cref@result}%
	\protected@xdef\cref@currentlabel{%
		[equation][\arabic{equation}][\cref@result]\p@equation\eq@number}}
		
\makeatother

\makeatletter
\let\cat@comma@active\@empty
\makeatother
\begin{document}

\title{Fluxonium Qubits in a Flip-Chip Package}

\author{Aaron Somoroff}
\affiliation{SEEQC, Inc., Elmsford, NY 10523}

\author{Patrick Truitt}
\affiliation{SEEQC, Inc., Elmsford, NY 10523}

\author{Adam Weis}
\affiliation{SEEQC, Inc., Elmsford, NY 10523}

\author{Jacob Bernhardt}
\affiliation{SEEQC, Inc., Elmsford, NY 10523}

\author{Daniel Yohannes}
\affiliation{SEEQC, Inc., Elmsford, NY 10523}

\author{Jason Walter}
\affiliation{SEEQC, Inc., Elmsford, NY 10523}

\author{Konstantin Kalashnikov}
\affiliation{SEEQC, Inc., Elmsford, NY 10523}

\author{Mario Renzullo}
\affiliation{SEEQC, Inc., Elmsford, NY 10523}

\author{Raymond A. Mencia}
\affiliation{Department of Physics, Joint Quantum Institute, and Quantum Materials Center, University of Maryland, College Park, MD 20742, USA}

\author{Maxim G. Vavilov}
\affiliation{Department of Physics and  Wisconsin Quantum Institute, University of Wisconsin - Madison, Madison, WI 53706, USA}

\author{Vladimir E. Manucharyan}
\affiliation{Department of Physics, Joint Quantum Institute, and Quantum Materials Center, University of Maryland, College Park, MD 20742, USA}
\affiliation{\'Ecole Polytechnique F\'ed\'erale de Lausanne, CH-1015 Lausanne, Switzerland}

\author{Igor V. Vernik}
\affiliation{SEEQC, Inc., Elmsford, NY 10523}

\author{Oleg A. Mukhanov}
\affiliation{SEEQC, Inc., Elmsford, NY 10523}

\date{\today}

\begin{abstract}

The strong anharmonicity and high coherence times inherent to fluxonium superconducting circuits are beneficial for quantum information processing. In addition to requiring high-quality physical qubits, a quantum processor needs to be assembled in a manner that minimizes crosstalk and decoherence. In this paper, we report work on fluxonium qubits packaged in a flip-chip architecture, where a classical control and readout chip is bump-bonded to the quantum chip, forming a multi-chip module (MCM). The modular approach allows for improved connectivity between the qubits and control/readout elements, and separate fabrication processes. We characterize the coherence properties of the individual fluxonium qubits, demonstrate high fidelity single-qubit gates with 6 ns microwave pulses (without DRAG), and identify the main decoherence mechanisms to improve on the reported results.

\end{abstract}

\maketitle

\section{Introduction}

Superconducting circuits are one of the leading quantum computing platforms due to the orders-of-magnitude growth in coherence times \cite{Devoret2013SuperconductingOutlook, Kjaergaard2020SuperconductingPlay} and a path to scalability derived from existing CMOS technologies \cite{Schoelkopf2008WiringSystems}. To date, superconducting quantum processors composed of many tens of qubits have been demonstrated \cite{Arute2019QuantumProcessor, Zhang2022High-fidelityQubits}. A fault-tolerant quantum computer capable of solving classically intractable problems would require millions of physical qubits \cite{Fowler2012SurfaceComputation}. This number can be lowered by improving the qubit performance. Nevertheless, the need for scaling up remains. 

A major obstacle to scaling is the overhead related to routing the multitude of lines for qubit control and readout in a planar architecture. The main challenges arise from signal crosstalk between qubits and the conflicting requirement of coupling high-Q qubits to lossier control and readout circuitry. Additionally, we expect that new qubit control methods will be necessary to realize large-scale quantum processors, since the power consumption and physical overhead of conventional analog microwave control is daunting. One example is single flux quantum (SFQ) technology, where the qubits are controlled using digital voltage pulses \cite{McDermott2018Quantum-classicalLogic, Leonard2019DigitalQubit}, which would greatly reduce the level of instrumentation overhead. 

A way to scale up from the planar level is to utilize multiple circuit layers; separating the quantum and classical circuitry \cite{Brecht2016MultilayerComputing,Foxen2018QubitInterconnects, Dunsworth2018ADevices}. One approach is the flip-chip package, where a carrier chip responsible for qubit control and readout is bump-bonded to the quantum chip (flip-chip), which is flipped such that it faces the carrier. The two chips are separated by vacuum or a chip interposer, and the qubit control lines and readout resonators couple capacitively or inductively to the qubit. 

The modular approach allows classical and quantum elements to be fabricated separately, thereby protecting the qubits from unnecessary exposure to additional fabrication processes. It also improves device yield, since defective qubits do not affect the functionality of the carrier chip, and vice versa. Furthermore, instead of routing signals from the perimeter of the chip, interconnects can be made through its plane, reducing the overlap between individual circuit elements \cite{Yost2020Solid-stateVias}. The flip-chip package is also optimal for digital qubit control using SFQ pulses because the quasiparticles \cite{Catelani2011RelaxationQubits, Glazman2021BogoliubovQubits} generated on the active carrier chip by the SFQ circuits are isolated from the quantum chip \cite{Liu2023SingleModuleb}. To date, flip-chip packaging has been explored on transmon \cite{Koch2007Charge-insensitiveBox} and capacitively-shunted flux 
 \cite{You2007Low-decoherenceQubit, Steffen2010High-coherenceQubit} qubits, demonstrating that the qubit performance is not limited by this architecture \cite{Rosenberg20173DQubits, Li2021Vacuum-gapTechnology, Kosen2022BuildingProcessor, Jurcevic2021DemonstrationSystem, Gold2021EntanglementDevice}. 

Over the past decade, the fluxonium circuit \cite{Manucharyan2009Fluxonium:Offsets} has emerged as a promising platform for superconducting quantum computing due to its extremely high coherence and strong anharmonicity \cite{Nguyen2019High-CoherenceQubit, Somoroff2023MillisecondQubit}. The latter gives fluxonium a clear advantage over the transmon since the concern of state leakage errors \cite{Chen2016MeasuringQubit} is mitigated when using arbitrarily short gates. Recent demonstrations of high fidelity single- and two-qubit gates \cite{Ficheux2021FastFluxoniums,Xiong2022ArbitraryShifts, Dogan2023Two-FluxoniumGate, Moskalenko2022HighCoupler, Bao2021Fluxonium:Operations, Ding2023High-FidelityCoupler} have given way to exploring methods of scaling up fluxonium-based systems \cite{Nguyen2022ScalableProcessorb}. One such method that has not been experimentally realized is embedding fluxonium in a multi-chip module (MCM). 

As an initial step, we present results on uncoupled fluxonium qubits embedded on a quantum chip (flip-chip), which is bump-bonded to a carrier chip for control and readout. We demonstrate comparable qubit performance in this new configuration to previous work on fluxoniums in 2D planar chips and 3D cavities \cite{Zhang2021UniversalQubit, Nguyen2019High-CoherenceQubit}. Additionally, we characterize the leading sources of decoherence and the single-qubit gate fidelities in this device.

\begin{figure*}
    \centering    \includegraphics[width=.89\linewidth]{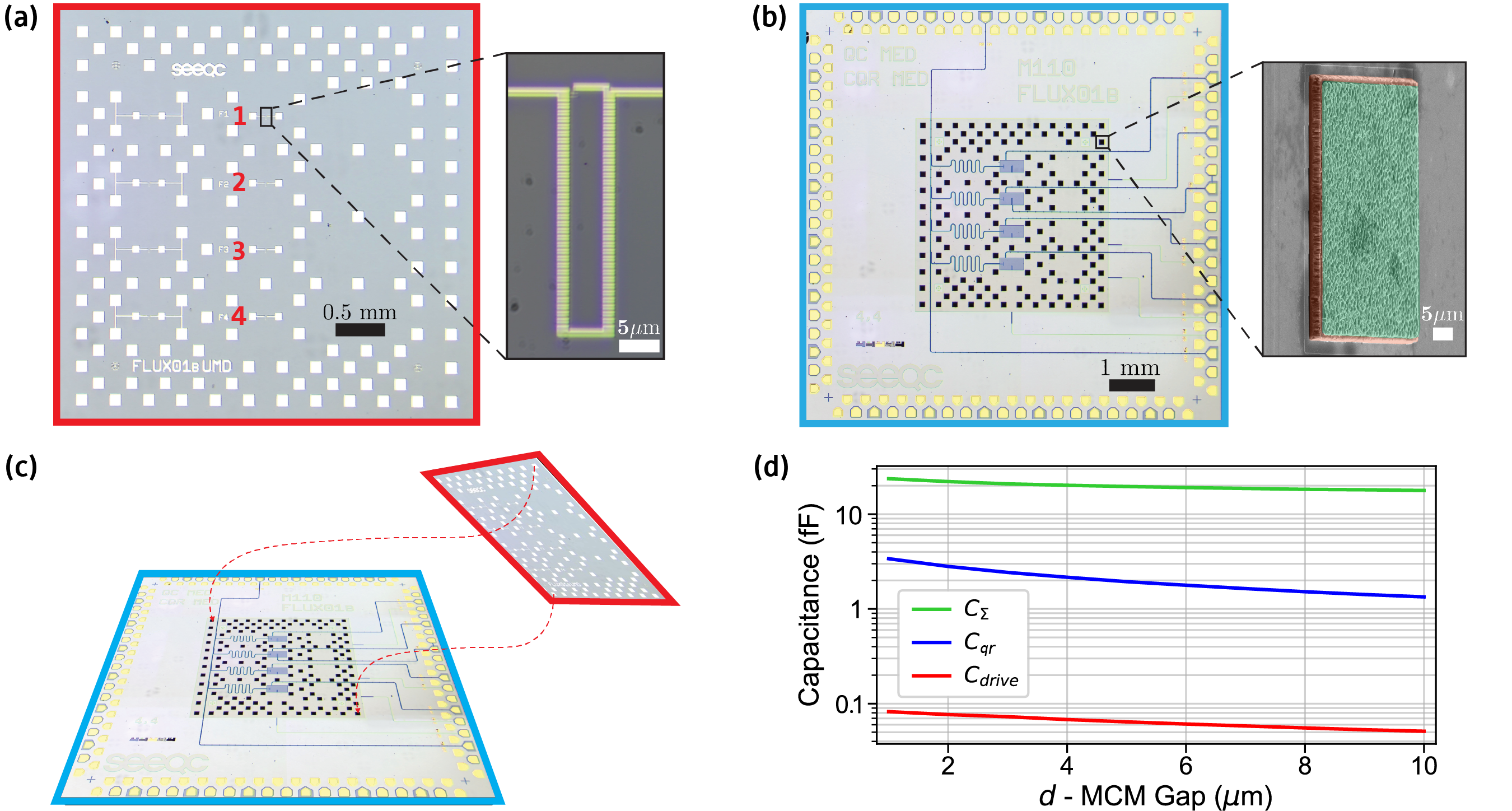}
	\caption{(a) The $5 \times 5 ~\text{mm}^2$ quantum chip (flip-chip), which houses four uncoupled fluxonium circuits with varying circuit parameters. The fluxoniums are labelled 1-4 (top to bottom). Aluminum pads are shown, which align with the 5 $\mu$m-high pillars on the carrier chip. Inset is a close-up of a fluxonium circuit. (b) The $10 \times 10 ~\mathrm{mm}^2$ carrier chip with classical circuitry for qubit control and readout. A single transmission line inductively couples to four $\lambda/4$ CPW resonators with fundamental frequencies in the $6-7~\mathrm{GHz}$ range, which capacitively couple to each fluxonium for readout. Additionally, there are four qubit drive lines and four DC-lines to flux bias the fluxonium loops. Inset is a false-color SEM image of one of the $120 \times 120$ $\mu \text{m}^2$-wide pillars. Orange(green) region is the 5(1) $\mu$m-thick aluminum(indium) layer. (c) Schematic of the configuration of flip-chip and carrier chip, which forms the multi-chip module. The flip-chip faces the carrier and the aluminum pads are aligned with the aluminum pillars on the carrier chip. The two chips are then bump-bonded together to construct the MCM (see Appendix \hyperref[app_fab]{E}). The pads on the carrier chip are wire-bonded to a printed circuit board (PCB), which connects to the coaxial cables in the dilution refrigerator (see Appendix \hyperref[app_exp]{F}). (d) Ansys Q3D simulation of total fluxonium shunt capacitance $C_{\Sigma}$, fluxonium-resonator coupling capacitance $C_{\mathrm{qr}}$, and fluxonium-drive line coupling capacitance $C_{\mathrm{drive}}$ versus MCM gap distance $d$.} 
	\label{fig:Fig1}
\end{figure*}

\section{Fluxonium Multi-Chip Module}

Our device consists of four uncoupled fluxonium circuits (labelled fluxoniums 1-4, see Figure \ref{fig:Fig1}) fabricated on a silicon chip, which is bump-bonded to the carrier chip with 5 $\mu$m  spacing. The passive carrier chip houses corresponding microwave control lines, DC lines for flux biasing, and $\lambda/4$ coplanar waveguide (CPW) resonators \cite{Goppl2008CoplanarElectrodynamics} for readout. The fluxonium circuit design is the same as in references \cite{Nguyen2019High-CoherenceQubit, Somoroff2023MillisecondQubit}, and can be further optimized for this new packaging method. All four circuits contain 138 large-area Josephson junctions in their superinductance arrays. The spectrum of the fluxonium circuits is varied by changing the Josephson energy $E_J$. Fluxoniums 1 and 2 were designed to have about twice larger $E_J$ than 3 and 4.
We targeted a qubit frequency of at least 1 GHz at the half flux sweet spot ($\Phi_{\text{ext}}/\Phi_0 = 0.5$) in order to reduce the thermal population in the $|1\rangle$ state and for a shorter gate duration. 

Images of the device are shown in Figure \ref{fig:Fig1}. The capacitance between the two ends of the fluxonium, as well as to the readout resonator and control line, was simulated with Ansys Q3D (see Figure \ref{fig:Fig1}d). Both ends of the fluxonium are ungrounded and the capacitance to the carrier groundplane contributes the majority of the fluxonium capacitance. To reduce this capacitance to meet our $E_C/h = 1 ~\mathrm{GHz}$ target, we removed the groundplane from a $300 \times 600$ $\mu$$\text{m}^2$ area of the carrier centered over the fluxonium. To account for the presence of the flip-chip on resonator frequencies, the CPWs that comprise the resonators were simulated with Sonnet \cite{SonnetSoftware}. 

\begin{figure}[h]
    \centering	\includegraphics[width=\linewidth]{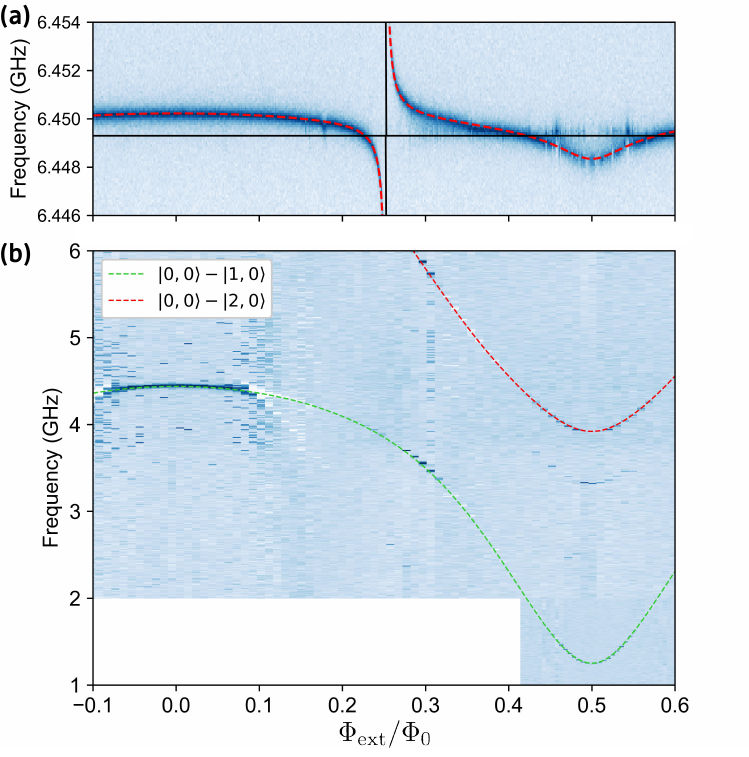}
	\caption{Spectroscopy of fluxonium 3 over approximately half a flux period. State $|j,k\rangle$ indicates $j$ excitations in the fluxonium circuit and $k$ excitations in the resonator. (a) One-tone spectroscopy of the readout resonator showing the avoided level crossing between the bare fluxonium circuit state $|2,0\rangle$ and the fundamental mode of the resonator $|0,1\rangle$. Dashed red curves are the theoretical fit to the coupled fluxonium-resonator Hamiltonian, yielding $\omega_r = 2\pi \times 6.4493 ~\mathrm{GHz}$ and $g = 2\pi \times 86 ~\mathrm{MHz}$. Horizontal and vertical solid black lines are the resonator and bare fluxonium $|0\rangle - |2\rangle$ transition, respectively, for $g = 0$. The level splitting between states $|2,0\rangle$ and $|0,1\rangle$ is given by $2g|\langle 2|\hat{n}|0\rangle| \approx 2\pi \times 31~\mathrm{MHz}$. (b) Two-tone spectroscopy of the fluxonium circuit depicting the qubit ($|0,0\rangle - |1,0\rangle$) and ground to second excited state ($|0,0\rangle - |2,0\rangle$) transitions. Dashed curves represent the same fit as in (a).
    }
	\label{fig:Fig2}
\end{figure}

The bare fluxonium Hamiltonian reads \cite{Manucharyan2009Fluxonium:Offsets}:
\begin{equation}
    \hat{H}_{\mathrm{fl}} = 4E_C \hat{n}^2 + \frac{1}{2} E_L\hat{\varphi}^2 - E_J\cos\left(\hat{\varphi} - 2\pi \frac{\Phi_{\mathrm{ext}}}{\Phi_0} \right),
    \label{eq;fluxoniumHamiltonian}
\end{equation} 
where the operators $\hat{n},~\hat{\varphi}$ represent the displacement charge across the capacitance in units of Cooper pairs and the supercondicting phase across the inductance, respectively. They obey $[\hat{\varphi}, \hat{n}] = i$. Additionally, the resonator at frequency $\omega_r$ has the familiar Hamiltonian term: $\hat{H}_{\mathrm{r}} = \hbar \omega_r(\hat{a}^{\dag}\hat{a} + 1/2)$, and the Hamiltonian term for fluxonium-resonator  coupling in the dispersive regime \cite{Wallraff2004StrongElectrodynamics, Blais2004CavityComputation} is given by: $\hat{H}_{\mathrm{c}} = -\hbar g\hat{n}(\hat{a}^{\dag} + \hat{a})$,
where $g$ is the fluxonium-resonator coupling rate. The spectroscopy data in Figure \ref{fig:Fig2} were fit to the numerical diagonalization of Hamiltonian: $\hat{H} = \hat{H}_{\mathrm{fl}} + \hat{H}_{\mathrm{r}} + \hat{H}_{\mathrm{c}} $
in the LC oscillator basis. 

We extensively characterized fluxoniums 3 and 4 since they met our 1 GHz qubit frequency target, and focus on fluxonium 3 in the main text. The parameters for both circuits are summarized in Table \ref{params}. We began our analysis of fluxonium 3 with one- and two-tone spectroscopy of the resonator and fluxonium circuit (Figures \ref{fig:Fig2}a and \ref{fig:Fig2}b, respectively). Dispersive readout of the qubit was performed using the corresponding resonator at frequency $\omega_r$, capacitively coupled to the fluxonium circuit with empirically determined coupling rate $g = 2\pi \times 86 ~\mathrm{MHz}$. The fit yields Josephson energy $E_J/h = 2.50 ~\mathrm{GHz}$, inductive energy $E_L/h = 1.14 ~\mathrm{GHz}$, and charging energy $E_C/h = 0.89 ~\mathrm{GHz}$. The qubit transition frequency at the sweet spot is $\omega_{01} = 2\pi \times 1.252 ~\mathrm{GHz}$. We used a Hilbert space of 25 fluxonium levels and 5 resonator levels in the fits.

The fitted values for $E_C$ and $g$ correspond to inter-chip distances $d$ of 2.3 and 1.3 $\mu$m, respectively. The discrepancy between the two distances can be explained by the fact that we did not account for lateral chip offsets nor chip tilt in the simulation, and $C_{\mathrm{qr}}$ depends more strongly on these variables than $C_{\Sigma}$ does. In general, a larger inter-chip distance will make the capacitances less sensitive to tilts, offsets, and variations in $d$ itself.

\begin{table}[h]
    \begin{tabular}{||c | c c||} 
     \hline
     Fluxonium & 3 & 4  \\ [0.5ex] 
     \hline\hline
     $E_J/h ~ (\mathrm{GHz})$ & 2.50 & 2.36  \\ 
     \hline
     $E_L/h ~ (\mathrm{GHz})$ & 1.14 & 1.14  \\ 
     \hline
     $E_C/h ~ (\mathrm{GHz})$ & 0.89 & 0.89  \\ 
     \hline
     $\omega_{01}/2\pi ~ (\mathrm{GHz})$ & 1.252 & 1.330  \\ 
     \hline
     $\omega_{12}/\omega_{01}$ & 2.14 & 1.99  \\ 
     \hline
     $T_1^{\mathrm{max}}~(\mu \text{s})$ & 77 & 58 \\ 
     \hline
     $T_2^{E, \mathrm{max}}~(\mu \text{s})$ & 38 & 33 \\ 
     \hline
     $\langle T_1 \rangle~(\mu \text{s})$ & 55.1 & 33.6 \\ 
     \hline
     $\langle T_2^{E} \rangle ~(\mu \text{s})$ & 27.5 & 23.2 \\ 
     \hline
     $ T_{\phi} ~(\mu \text{s})$ & 36.6 & 35.4 \\ 
     \hline
     $\omega_r/2\pi~(\mathrm{GHz})$ & 6.4493 & 6.1391 \\ 
     \hline
     $g/2\pi~(\mathrm{MHz})$ & 86 & 85 \\ 
     \hline
     $\chi_{01}/2\pi~(\mathrm{MHz})$ & 1.39 & 0.63 \\ 
     \hline
     $\kappa/2\pi~(\mathrm{MHz})$ & 0.391 & 0.269 \\ 
     \hline
     $\tan\delta_C~(\mathrm{\times 10^{-6}})$ & 1.6 & 2.0 \\ 
     \hline
     $n_{\mathrm{th}}~(\times 10^{-2})$ & 1.2 & 2.0 \\ 
     \hline
    \end{tabular}
    \caption{Summary of relevant parameters for each fluxonium circuit under study at half flux quantum $\Phi_{\mathrm{ext}}/\Phi_0 = 0.5$.}
    \label{params}
\end{table}

\section{Qubit Performance}

\begin{figure}[h]
    \centering
    \includegraphics[width=0.98\linewidth]{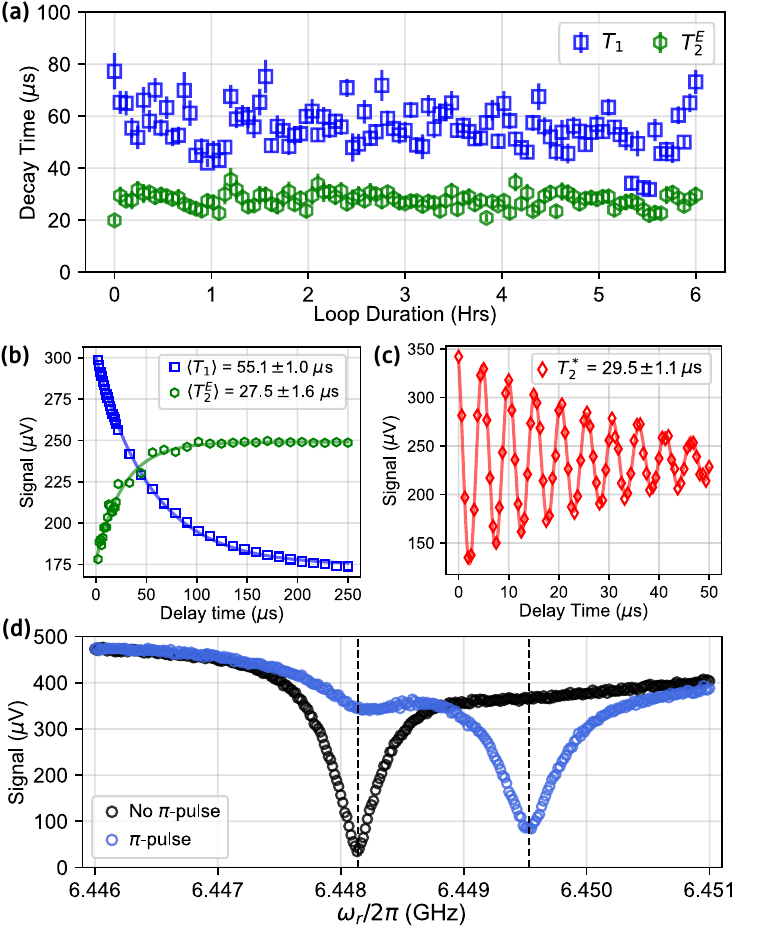}
	\caption{Measurements of the fluxonium 3 qubit transition at $\Phi_{\mathrm{ext}}/\Phi_0 = 0.5$. (a) Interleaved $T_1$ and $T_2^E$ loop taken over 6 hours. (b) $T_1$ and $T_2^E$ signals averaged over all 101 measurements in loop, fitted to decaying exponentials. The fitted decay time constants are the mean relaxation time $\langle T_1 \rangle = 55.1 \pm 1.0$  $\mu$s and Hahn-Echo coherence time $\langle T_2^E \rangle = 27.5 \pm 1.6$  $\mu$s. (c) A single Ramsey measurement fitted to an exponentially decaying cosine function, yielding $T_2^* = 29.5 \pm 1.1$  $\mu$s. (d) One-tone spectroscopy of the readout resonator with (light blue data) and without (black data) a $\pi$-pulse applied to the qubit to flip the qubit state. Dashed vertical lines denote the center frequency of each fitted Lorentzian function. The difference of the two frequencies is the qubit dispersive shift $\chi_{01} = 2\pi \times 1.39~\mathrm{MHz}$.}
	\label{fig:Fig3}
\end{figure}

We measured the energy relaxation time $T_1$ by applying a $\pi$-pulse to the qubit and varying the readout delay. At maximum, $T_1 = 77.3 \pm 7.0 $ $\mu$s (see Figure \ref{fig:Fig3}a). The effective dielectric loss tangent $\tan\delta_C$ of the total circuit capacitance $C_{\Sigma} = e^2/2E_C$ corresponding to this maximal $T_1$ at qubit frequency 1.252 GHz is approximately $1.6 \times 10^{-6}$. This value is similar to those reported in 3D fluxonium devices fabricated with the same materials and process \cite{Nguyen2019High-CoherenceQubit}, as well as in planar 2D architectures \cite{Zhang2021UniversalQubit}.
The effect of the MCM configuration on $T_1$ may vary, however, depending on the carrier groundplane geometry and the inter-chip distance. By optimizing the materials and fabrication process \cite{Place2021NewMilliseconds, Somoroff2023MillisecondQubit}, a reduction in $\tan\delta_C$ by a factor of two is within reasonable expectation. 

For calculating $\tan\delta_C$, it follows from Fermi's Golden Rule that the decay rate between eigenstates $|j\rangle,~|k\rangle$ due to dielectric loss \cite{Martinis2005DecoherenceLoss} in the capacitor at a finite temperature $T$ reads \cite{Nguyen2019High-CoherenceQubit, Zhang2021UniversalQubit, Masluk2012ReducingAtom}:

\begin{equation}
    \label{diel}
    \Gamma_{jk}^{\mathrm{diel}} =  \frac{8 E_C}{\hbar}|\hat{n}_{jk}|^2 \tan \delta_C \left[ 1 + \coth{\left(\frac{\hbar \omega_{jk}}{2k_B T} \right)} \right ].
\end{equation}
Here, $\hat{n}_{jk} \equiv \langle j |\hat{n}| k \rangle$ is the fluxonium charge matrix element between states $|j\rangle$ and $|k\rangle$ in units of Cooper pairs. The loss tangent $\tan\delta_C$ can be thought of as the inverse quality factor of the dielectric. We used Equation \ref{diel} to calculate corresponding loss tangents for fluxoniums 3 and 4, given their maximal $T_1$ values at the sweet spot. We assume a qubit temperature of $T = 20~\mathrm{mK}$, the approximate base temperature of the dilution refrigerator. Note that due to $\omega_{01} > 1$ GHz in our experiments, the extracted $\tan\delta_C$ depends weakly on the temperature. For example, $T = 60$ mK would correspond to $\tan\delta_C = 1.0 \times 10^{-6}$. Since $\tan\delta_C$ can depend weakly on frequency, the reported values for fluxoniums 3 and 4 should be interpreted as the value at their qubit frequencies (see Table \ref{params}).

The coherence time $T_2$ was measured via Ramsey and Hahn-Echo experiments. In the Ramsey method \cite{Ramsey1950AFields}, two $\pi/2$-pulses are applied to the qubit, and the time delay between them is swept. Readout immediately follows the second $\pi/2$-pulse. This pulse sequence yields Ramsey fringes with an exponentially decaying envelope with decay constant $T_2^*$, the Ramsey coherence time. In the Hahn-Echo sequence \cite{Hahn1950SpinEchoes}, a $\pi$-pulse is inserted between the two $\pi/2$-pulses. This refocusing $\pi$-pulse reverses any phase accumulation due to low-frequency drifts during the first half of the sequence \cite{Martinis2003DecoherenceNoise}. By increasing the number of refocusing $\pi$-pulses, known as a Carr-Purcell-Meiboom-Gill (CPMG) sequence \cite{Meiboom1958ModifiedTimes}, the qubit will become more sensitive to noise at successively higher frequencies \cite{Bylander2011NoiseQubit}. At maximum, we measured $T_2^* = 29.5 \pm 1.1$ $\mu$s and Hahn-Echo coherence time $T_2^E = 36.3 \pm 4.0$ $\mu$s (Figures \ref{fig:Fig3}a, c). 



We performed an interleaved $T_1$, $T_2^E$ loop to obtain statistics on the relaxation and coherence times of fluxonium 3, and found $\langle T_1 \rangle = 55.1 \pm 1.0$ $\mu$s and $\langle T_2^E \rangle = 27.5 \pm 1.6$ $\mu$s. $\langle T_2^E \rangle < \langle T_1 \rangle$ reveals significant pure dephasing of the qubit. We defined the dephasing time as $T_{\phi} = 1/(1/\langle T_2^E \rangle - 1/2\langle T_1 \rangle)$, yielding $T_{\phi} = 36.6 \pm 2.5$ $\mu$s. The exponentially decaying Hahn-Echo and Ramsey signals (Figures \ref{fig:Fig3}b and c, respectfully) rule out $1/f$ flux noise \cite{Yoshihara2006DecoherenceNoise,Kumar2016OriginDevices} as the predominant dephasing source. When we applied multiple refocusing pulses in a CPMG sequence, the measured decay time did not increase, indicating dephasing due to white noise (see Appendix \hyperref[app_coh]{B}). It is therefore most likely that the pure dephasing of the qubit is caused by thermal photons in the resonator \cite{Bertet2005DephasingNoise}. Based on the measured qubit dispersive shift $\chi_{01}$ and resonator linewidth $\kappa$ (see Figure \ref{fig:Fig3}d and Table \ref{params}), $T_{\phi} = 36.6 \pm 2.5$ $\mu$s corresponds to an average thermal photon number of $n_{\textup{th}} = (1.2 \pm 0.1) \times 10^{-2}$ and an effective resonator temperature of $T_{\mathrm{res}} = 70 \pm 1~\mathrm{mK}$ (see Appendix \hyperref[app_ro]{A}). Our measured values for $n_{\textup{th}}$ and $T_{\mathrm{res}}$ agree with those reported in transmon and flux qubits \cite{Wang2019CavityQubits, Yeh2017MicrowaveMK, Yan2016TheReproducibility}. 


\begin{figure}[h]
    \centering
    \includegraphics[width=1.0\linewidth]{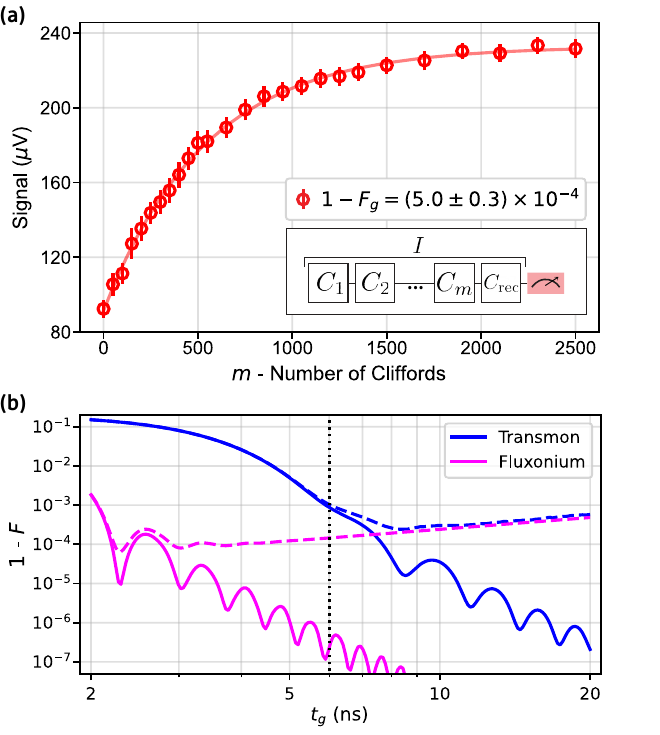}
	\caption{(a) Single-qubit randomized benchmarking on fluxonium 3. The red data points are the result of averaging over 32 randomizations of the RB sequence. The fit yields an error rate per Clifford of $r_{\mathrm{cliff}} = (9.2 \pm 0.6) \times 10^{-4}$. 
    (b) Simulated $X/2$ gate error rate $1-F$ versus gate duration $t_g$ for fluxonium 3 and a transmon (see main text) with (dashed curves) and without (solid curves) decoherence. When including decoherence, we used the measured $T_1$ and $T_{\phi}$ for fluxonium 3, given in Table \ref{params}. Basic (no DRAG), cosine-envelope pulses were used in the simulation. Dashed black line corresponds to the gate duration ($6~\text{ns}$) used in our experiments.}
	\label{fig:Fig4}
\end{figure}

We completed our study of the fluxonium MCM by characterizing the single-qubit gates using randomized benchmarking (RB) \cite{Magesan2011ScalableProcesses,  Magesan2012EfficientBenchmarking}. The Clifford gates were generated by the gate set $\{I, X, Y, \pm X/2, \pm Y/2 \}$; we defined the identity gate $I$ as no pulse. An RB sequence is shown as an inset in Figure \ref{fig:Fig4}a. The physical pulses making up the $\pm X/2, \pm Y/2$ gates were cosine-envelope microwave pulses with total gate duration $t_g = 6~\mathrm{ns}$. The $X,~Y$ gates were composed of two concatenated $X/2, Y/2$ gates. 


Figure \ref{fig:Fig4}a shows the results of the RB experiment on fluxonium 3, along with the fit, which yields an average Clifford gate error rate of $r_{\mathrm{cliff}} = (9.2 \pm 0.6) \times 10^{-4}$ and an average gate error rate of $r_g = r_{\mathrm{cliff}}/1.833 = (5.0 \pm 0.3) \times 10^{-4}$. Given $T_2^E \approx 30$ $\mu$s, and the average Clifford gate duration of $t_g^{\mathrm{cliff}} = $ 13 ns,
we estimate the coherence limit for the Clifford gate error rate to be: $t_g^{\mathrm{cliff}}/T_2^E \approx 4.3 \times 10^{-4}$. Comparing this estimate to the measured $r_{\mathrm{cliff}}$, we conclude that at least half of our gate error can be attributed to imperfect calibration. Additional pulse optimization, such as DRAG \cite{Motzoi2009SimpleQubits}, can help approach the coherence limit. More details on the microwave pulses and measurement of the individual gate error rates from interleaved RB \cite{Magesan2012EfficientBenchmarking} are found in Appendix \hyperref[app_gate]{C}.

In Figure \ref{fig:Fig4}b, we simulated the gate error rate $1-F$ versus total gate duration $t_g$ of a $X/2$ gate for fluxonium 3 (anharmonicity $\alpha = 1.42~\text{GHz}$) and a transmon ($E_J/h = 15~\text{GHz}$, $E_C/h = 0.3~\text{GHz}$, $\alpha = -0.345~\text{GHz}$). The anharmonicity is defined as $\alpha = (\omega_{12} - \omega_{01})/2\pi$. The solid curves are the gate error rate with no decoherence in the system, where state leakage outside of the computational subspace is the dominant error source. The dashed curves include relaxation and dephasing using the measured $T_1$ and $T_{\phi}$ of fluxonium 3. Details of the simulation are provided in Appendix \hyperref[app_gate]{C}. 

For $t_g = 6~\mathrm{ns}$ in our experiments, the simulated coherence limit on the $X/2$ gate error rate is $1.5 \times 10^{-4}$. Given the measured $X/2$ error rate (see Table \ref{gate}) of $3.1 \times 10^{-4}$, we estimate that 48\% of the gate error is due to decoherence. The discrepancy between the measured gate error rate and the coherence limit can be attributed to the 1 GSa/s sampling rate of the AWG. In the simulation, we assume a nearly continuous pulse envelope, while in practice this cannot be achieved as we approach $t_g = 1 ~\mathrm{ns}$.

Crucially, Figure \ref{fig:Fig4}b shows that it is impossible to reach the same gate error rates using basic (no DRAG) $6~\mathrm{ns}$ microwave pulses on a typical transmon qubit, even without decoherence. Due to fluxonium's higher anharmonicity, such a short gate time is accessible. Only at a gate duration of around 3 ns does the error rate become dominated by state leakage. Consequently, we could achieve a state-of-the-art gate fidelity in this device, despite a lower coherence time.


\section{Conclusion}

In conclusion, we have packaged fluxonium qubits in a multi-chip module, and demonstrated performance on par with the previously cited experiments in 2D and 3D packages. An effective resonator temperature of 70 mK and average thermal photon number $n_{\mathrm{th}} = 1.2 \times 10^{-2}$ is sufficient to explain the measured $T_{\phi}$. By optimizing the readout line filtering and attenuation, along with reducing $\chi_{01}/\kappa$, we expect the coherence time to approach the $2T_1$ limit. 

Despite the high level of pure dephasing, a single-qubit gate fidelity greater than 0.999 was possible by using 6 ns pulses. A conservative goal of doubling the reported $T_{\phi}$, without any improvement of $T_1$, would push the coherence limit on the gate fidelity to above 0.9999. Fast gates, calibrated in a straightforward manner, are a key advantage of fluxonium, where the anharmonicity is about an order of magnitude larger than in transmons. Our results on the first fluxonium MCM may be useful for scaling the next generation of fluxonium-based quantum processors. \\

\noindent We acknowledge contributions from SEEQC foundry engineers John Vivalda and Asa Chambal-Jacobs. 

\section*{Appendix}

\subsection{Readout}
\label{app_ro}
\subsubsection{Qubit-Resonator Interaction}
\label{ro_characterize}

Each fluxonium circuit is capacitively coupled to a $\lambda/4$ CPW resonator, which is housed on the carrier chip. The fluxonium-resonator coupling rate $g$ was found empirically by fitting the spectroscopy data in Figure \ref{fig:Fig2}. The pull on the resonator frequency due to bare fluxonium state $|j\rangle$, known as the dispersive shift of state $|j\rangle$, is given by \cite{Lin2018DemonstrationDecay}:

\begin{equation}
    \begin{aligned}
        \chi_j = g^2 \sum_{ k } |\hat{n}_{jk}|^2 \frac{2\omega_{jk}}{\omega_{jk}^2 - \omega_r^2}.
    \end{aligned}
    \label{eq:chi}
\end{equation}

\begin{figure}[h]
    \centering	\includegraphics[width=\linewidth]{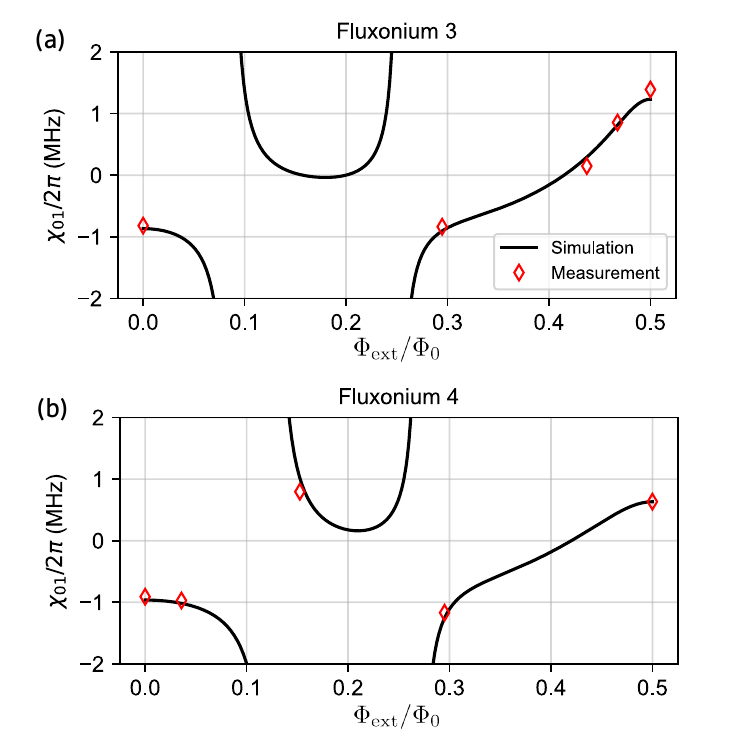}
	\caption{Simulated qubit dispersive shift $\chi_{01}$ for fluxoniums 3 (a) and 4 (b) using Equation \ref{eq:chi} and the parameters in Table \ref{params}. We include measured values of $\chi_{01}$ at various external flux biases.}
	\label{fig:chi_sim}
\end{figure}

The dispersive shift of the transition between states $|i\rangle$ and $|j\rangle$ is then $\chi_{ij} = \chi_{i} - \chi_{j}$. Crucially, Equation \ref{eq:chi} shows that the qubit transition can have a large dispersive shift due to the contributions of the higher levels where $\omega_{jk} \simeq \omega_r$, despite the large detuning between $\omega_{01}$ and $\omega_r$. Figure \ref{fig:chi_sim} uses Equation \ref{eq:chi} to plot $\chi_{01}$ across half a flux period for fluxoniums 3 and 4. Red diamonds are measured $\chi_{01}$ at various external flux biases, showing excellent agreement between theory and experiment. The simulation uses 25 fluxonium levels, 20 of which are used in the summation to find $\chi_{01}$.

In our system, the fluxonium-resonator coupling rate $g$ in Equation \ref{eq:chi} is given by:

\begin{equation}
    \begin{aligned}
       g = \frac{1}{2}\frac{C_{\text{qr}}}{C_{\Sigma}C_{r}}\frac{1}{\sqrt{\zeta_q \zeta_r}},
    \end{aligned}
    \label{eq:g}
\end{equation}
where $\zeta_q = \frac{R_q}{2\pi}\sqrt{\frac{8E_C}{E_L+E_J}}$ and $\zeta_r = \sqrt{\frac{L_r}{C_r}}$ are the fluxonium and resonator characteristic impedances, respectfully. $R_q = h/(2e)^2$ is the resistance quantum. 

\subsubsection{Thermal Photon Dephasing}
\label{cav_photon}

The qubit dephasing rate due to thermal photons in the cavity is given by:
\begin{equation}
    \Gamma_{\phi}^{\textup{th}} = \frac{n_{\textup{th}}\kappa \chi_{01}^2}{\kappa^2+\chi_{01}^2}
\end{equation}
in the low photon number $n_{\textup{th}}$ limit \cite{Clerk2007UsingOscillator,Wang2019CavityQubits}. Plugging in the values of $\kappa$, $\chi_{01}$, and $ T_{\phi}$ for each fluxonium (see Table \ref{params}), we find the average thermal photon numbers $n_{\textup{th}} = (1.2 \pm 0.1) \times 10^{-2}$ and $(2.0 \pm 0.2) \times 10^{-2}$ for fluxoniums 3 and 4, respectively. These values correspond to resonator temperatures of $70 \pm 1~\mathrm{mK}$ and $75 \pm 2~\mathrm{mK}$ for fluxoniums 3 and 4, respectively. 

\subsection{Coherence Analysis}
\label{app_coh}

\subsubsection{$1/f$ Flux Noise}
\label{fluxnoise}

$1/f$ flux noise is known to be a leading cause of pure dephasing in flux-sensitive qubits, with a noise spectral density \cite{Martinis2003DecoherenceNoise,Yoshihara2006DecoherenceNoise,Koch2007ModelQubits,Kumar2016OriginDevices,Yan2016TheReproducibility}:

\begin{equation}
    S_{\Phi}(\omega) = 2\pi \frac{A_{\Phi}^2}{\omega},
\end{equation}
where $A_{\Phi}$ is the flux noise amplitude at 1 Hz. The dephasing rate due to first-order $1/f$ flux noise is given by:
\begin{equation}
    \Gamma_{\Phi} = A_{\Phi} \sqrt{\mathrm{ln}2}\frac{\partial \omega}{\partial \Phi_{\mathrm{ext}}}.
\end{equation}

\begin{figure}[h]
	\centering
	\includegraphics[width=.95\linewidth]{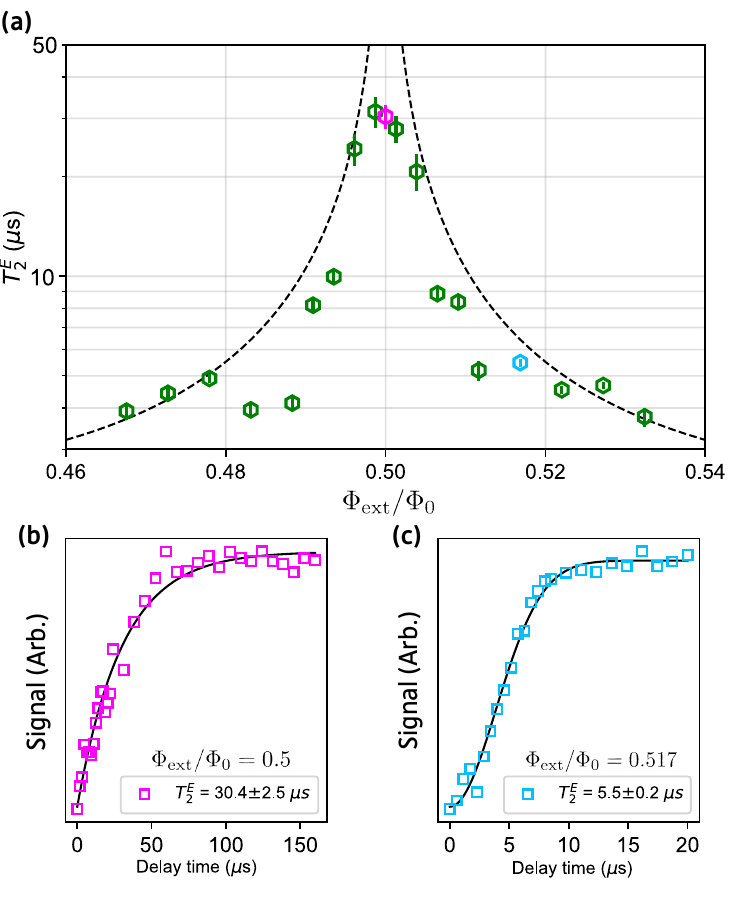}
	\caption{(a) $T_2^E$ versus external magnetic flux $\Phi_{\mathrm{ext}}$. 
     The data are fit to a decaying Gaussian function off sweet spot and to a decaying exponential at the sweet spot. Dashed black curve represents the simulated values for first-order flux noise-induced dephasing time $T_{\phi}^{\Phi} = 1/\Gamma_{\Phi}$ for flux noise amplitude $A_{\Phi} = 5.5 ~\mu \Phi_0/\sqrt{\mathrm{Hz}}$ at 1 Hz. At the sweet spot, the first-order limit diverges, and coherence is mainly limited by energy relaxation and uncorrelated (white) noise. (b) Measurement trace at the sweet spot. The data are fit to a decaying exponential and the decay constant corresponds to the magenta point in (a). (c) Measurement trace at a flux bias point off the sweet spot. The data are fit to a decaying Gaussian and the decay constant corresponds to the blue point in (a).}
	\label{fig:t2_flux}
\end{figure} 

This leads to a $T_2^E$ measurement signal decaying with a Gaussian envelope, with decay constant $T_{\phi}^{\Phi}=1/\Gamma_{\Phi}$. This is the case when fluxonium is flux biased off of the sweet spot. On the sweet spot, the first-order sensitivity to $1/f$ flux noise vanishes, and we observe an exponentially decaying echo signal \cite{Nguyen2019High-CoherenceQubit}. To determine $A_{\Phi}$ in our experimental setup, we measured $T_2^E$ at various $\Phi_{\mathrm{ext}}$ in fluxonium 3 (see Figure \ref{fig:t2_flux}). The observed dependence of $T_2^E$ versus external flux bias corresponds to $A_{\Phi} = 5.5 $ $\mu$$\Phi_0/\sqrt{\mathrm{Hz}}$, which is similar to the previously cited work. 

\subsubsection{CPMG}

\begin{figure}[h]
	\centering
	\includegraphics[width=.95\linewidth]{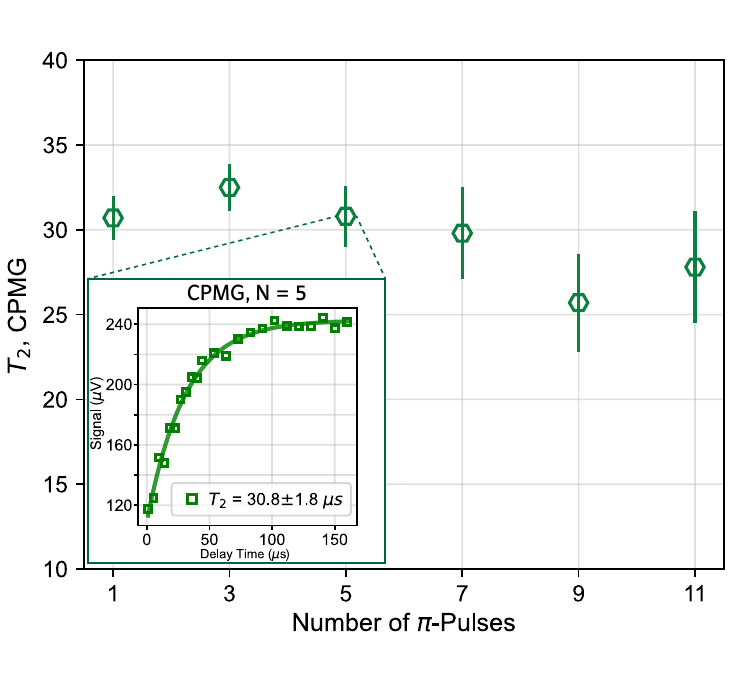}
	\caption{CPMG experiment on fluxonium 3 where we sweep the number of refocusing $\pi$-pulses $N$ from 1 to 11. The resulting measurement traces are then fitted to a decaying exponential function and the decay constant is plotted. We find that there is no significant correlation between number of pulses and $T_2$ in this range. Inset shows the measurement trace and fit for 5 intermediate $\pi$-pulses.}
	\label{fig:t2_cpmg}
\end{figure}

By increasing the number of refocusing $\pi$-pulses $N$ in a Hahn-Echo sequence (where $N = 1$), the qubit will be more sensitive to noise at successively higher frequencies \cite{Martinis2003DecoherenceNoise, Cywinski2008HowQubits}. This is known as a Carr-Purcell-Meiboom-Gill (CPMG) sequence \cite{Meiboom1958ModifiedTimes}, where the additional refocusing $\pi$-pulses act as a bandpass filter centered at a frequency determined by $N$ and the free evolution time \cite{Bylander2011NoiseQubit, Yan2016TheReproducibility}. If the noise spectrum is white within the region we sweep our filter, we expect the measured decay time to remain constant with an increasing number of $\pi$-pulses. If the noise spectral density $\propto 1/f$, we expect $T_2$ to rise for $N>1$. Given the results of our CPMG experiment shown in Figure \ref{fig:t2_cpmg}, we concluded that thermal resonator photons are the main dephasing source at the sweet spot.

\subsection{Single-Qubit Gates} 
\label{app_gate}

\subsubsection{Microwave Pulses}

The $X/2$ and $Y/2$ pulses were synthesized with a Keysight M3202 AWG and a Rohde \& Schwarz SGS100A Vector Modulation RF source for low frequency IQ modulation. We used cosine-envelope pulses with an envelope function:

\begin{equation}
    f(t) = \frac{\epsilon}{2}[1 - \cos(2\pi t/t_g)],
    \label{eq:envelope}
\end{equation}
where $\epsilon$ is the drive amplitude and $t_g$ is the total gate duration. We set $t_g = 6~\mathrm{ns}$ in our experiments. The cosine envelope is chosen over Gaussian to avoid truncation at the edges. The $X$ ($Y$) pulses consist of pairs of concatenated $X/2$ ($Y/2)$ pulses. We tuned the qubit drive frequency $\omega_d$ by taking Ramsey measurements with varying $\omega_d$. After obtaining the $\omega_d = \omega_{01}$, we performed a pulse train measurement to determine the optimal pulse amplitude. The pulse train consisted of sweeping the amplitude for a train of 4, 16, 36, 64, 100, and 144 pulses. The resulting trace of each measurement is effectively a Rabi oscillation, where the Rabi frequency increases with the number of pulses in the train. The amplitude corresponding to the same phase in the Rabi oscillations across all the trains is the optimal pulse amplitude. After performing initial randomized benchmarking experiments, the pulse amplitude was further fine-tuned using the ORBIT procedure \cite{Kelly2014OptimalBenchmarking}.

\subsubsection{Single-Qubit Randomized Benchmarking}

We benchmarked the single-qubit gates in fluxoniums 3 and 4 using randomized benchmarking (RB) \cite{Magesan2011ScalableProcesses}. In the RB sequence, $m$ randomly chosen Clifford gates are applied to the qubit before applying a single recovery gate, such that the entire sequence is tantamount to an identity operation. In practice, due to decoherence and sub-optimal pulse parameters, the RB signal decays with the number of Clifford gates in the sequence $m$ as:
\begin{equation}
    A + Bp^m,
    \label{rb_func}
\end{equation}
where $p$ is the depolarization parameter, and $A, B$ are constants. We extract $p$ by fitting the measured RB signal to Equation \ref{rb_func} (see Figure \ref{fig:Fig4}). The average error rate of a single-qubit Clifford operation $r_{\textup{cliff}}$ is then given by:
\begin{equation}
    r_{\textup{cliff}} = (1-p)/2.
\end{equation}
Since each Clifford operation is composed on average of 1.833 physical gates (we don't count the identity gate), the average physical gate fidelity is: 

\begin{equation}
    F = 1 - r_{\textup{cliff}}/1.833.
\end{equation}
We estimated the lower limit on the gate error rate by taking $t_g/T_2^E$. 

To determine the individual gate fidelities we performed interleaved randomized benchmarking \cite{Magesan2012EfficientBenchmarking}. The error rates of each gate, $1-F_g$, are given in Table \ref{gate}. In the interleaved RB sequence, a given gate is interleaved between each Clifford operation. The resulting curve follows the same decay profile as the standard RB, but with a depolarization parameter $p_{g}$. The gate error is then given by:

\begin{equation}
    r_{g} =(1-p_{g}/p)/2 = 1-F_{g},
\end{equation}
where $p$ is the depolarization parameter obtained from the initial RB and $F_g$ is the fidelity of the interleaved gate. 

\begin{table}[h]
    
    \begin{tabular}{||c | c | c | c | c | c | c||} 
     \hline
     Gate & X & X/2 & -X/2 & Y & Y/2 & -Y/2  \\ [0.5ex] 
     \hline\hline
     $1 - F_g ~(\times 10^{-4})$ & 8.1 & 3.1 & 2.6 & 6.1 & 2.1 & 5.9  \\ 
     \hline
     
    \end{tabular}
    \caption{Gate error rates $1 - F_g$ of each Clifford-generating gate obtained from interleaved RB for fluxonium 3. The relative error of each error rate is approximately 10\%.}
    \label{gate}
\end{table}

\subsubsection{Gate Simulation}

The simulation results shown in Figure \ref{fig:Fig4}b were obtained as follows. We truncated the fluxonium and transmon Hamiltonians to only consider the five lowest energy eigenstates. We then define a pulse with driving frequency $\omega_d = \omega_{01}$, and envelope given by Equation \ref{eq:envelope}, and compute the evolution operator $U$ governing the time dynamics of the circuit when the pulse is applied. Sweeping the gate time $t_g$, we optimize the amplitude $\epsilon$ and drive frequency $\omega_d$ of the pulse using Sequential Least Squares Programming to minimize the gate error rate $1-F$, where the gate fidelity $F$ is defined as \cite{Pedersen2007FidelityOperations}:

\begin{equation}
    F = \frac{1}{6} \left[ \mathrm{Tr}(U^{\dag} U) + |\mathrm{Tr}(U^{\dag} U_{\mathrm{Ideal}})|^2 \right],
\end{equation}
where $U_{\mathrm{Ideal}} = R_X (\pi/2)$, or a $X/2$ gate. Note that $U$ is non-unitary due to driving of higher energy circuit transitions. 

To simulate the gate error with decoherence, we used the Lindblad master equation to model the evolution of the density matrix $\rho$ with the optimal pulse parameters obtained from the simulation without decoherence. We included two collapse operators representing relaxation from qubit excited to ground state, and pure dephasing between the qubit states, following the procedure presented in \cite{Nesterov2021ProposalTransition}.


\subsection{Flux Bias Crosstalk}
\label{app_xtalk}
\begin{figure}[h]
    \centering
    \includegraphics[width=.95\linewidth]{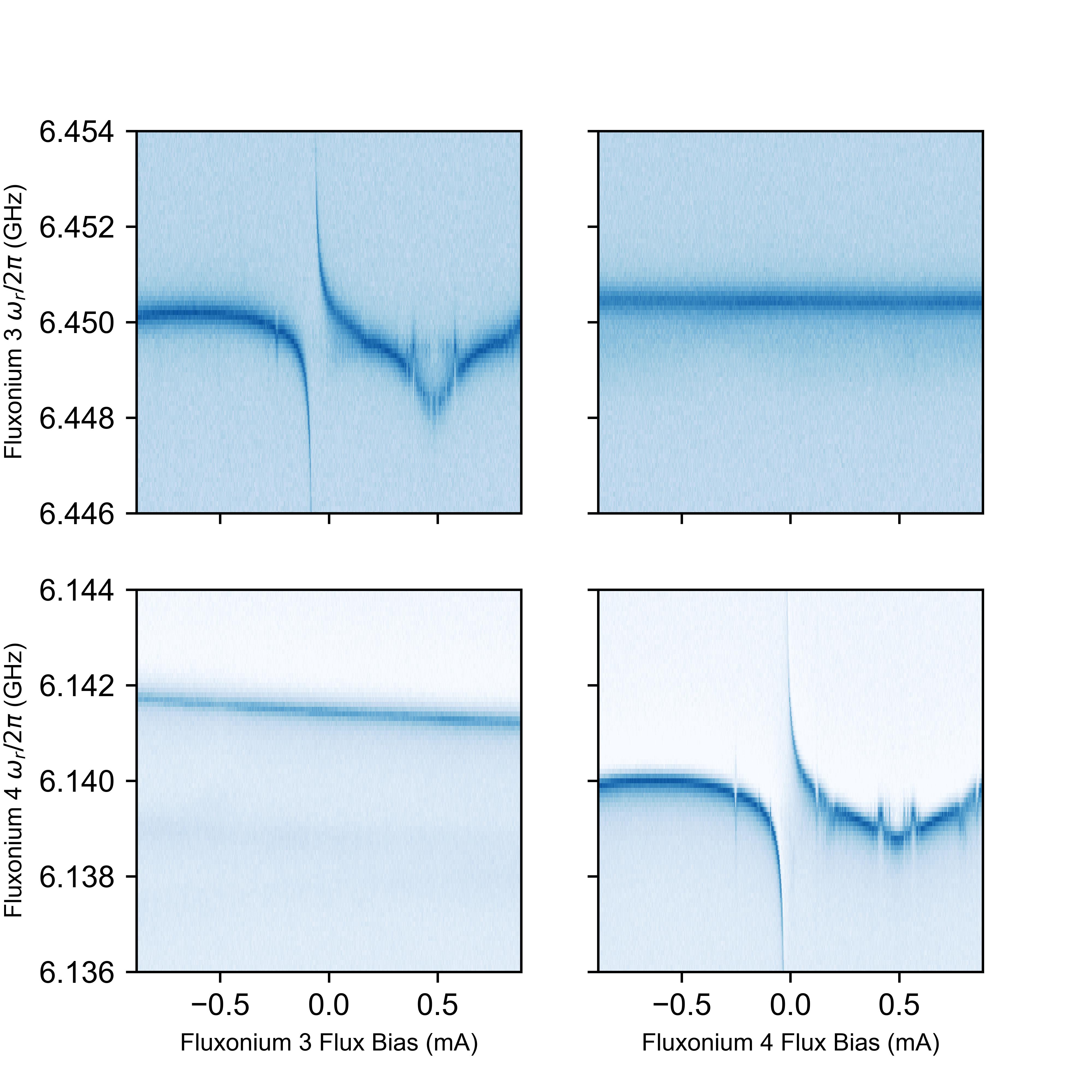}
	\caption{Flux crosstalk matrix between fluxoniums 3 and 4. Spectroscopy of fluxoniums 3 and 4 resonators is performed while sweeping the current applied to each bias line on the carrier chip.}
    \label{fig:xtalk}
\end{figure}

We performed resonator spectroscopy on both fluxoniums 3 and 4 while tuning their respective external flux biases. We constructed a flux crosstalk matrix (Figure \ref{fig:xtalk}) to quantify the mutual inductance between the flux bias line of fluxonium 3 and fluxonium 4's superinductance loop, and vice versa. The mutual inductance between the fluxonium circuit loop and flux bias line for fluxoniums 3 and 4 is 2.27 $\mathrm{mA}/\Phi_0$ and 2.23 $\mathrm{mA}/\Phi_0$, respectively. This is determined using the diagonal elements of the flux crosstalk matrix. The mutual inductance between fluxonium 4 and fluxonium 3's bias line is 430 $\mathrm{mA}/\Phi_0$. Fluxonium 3's resonator frequency did not shift over the current range of fluxonium 4's bias line, therefore this mutual inductance is negligible.

\subsection{Fabrication}
\label{app_fab}

\subsubsection{Quantum Chip}

We fabricated the device on a high resistivity silicon substrate. \textit{Cleaning:} The chip is prepared by sonicating it in acetone, then isopropyl alcohol (IPA) for 3 minutes each. \textit{Electron Beam Resist Application:} 1 drop of MMA EL-13 electron beam resist is applied to the chip, then spun at 5000 RPM for 1 minute. The resist is then baked on a hotplate for 1 minute at $180^{\circ}$C. A second layer of resist is then applied: 1 drop of 950 PMMA A3 electron beam resist spun at 4000 RPM for 1 minute before baking at $180^{\circ}$C for 30 minutes. \textit{Electron Beam Lithography:} The circuit is written with a 100 kV Elionix Electron Beam Lithography system, using a beam current of 1 nA. \textit{Resist Development:} Mask is developed for 2 minutes in a 3:1 IPA:DI solution at $6^{\circ}$C. The chip is lightly shaken back and forth by hand at around 1-2 Hz while in the developer. \textit{Metal Deposition:} Chip is loaded into a Plassys deposition system and the loadlock is pumped on for 20 hours until the pressure reaches $1.3 \times 10^{-7}$ mBar prior to deposition. The deposition is comprised of the following steps: 1. 20 second Ar etch at each deposition angle ($\pm 23.83^{\circ}$) 2. Deposit Ti into the chamber at 0.1 nm/s for 2 minutes 3. First Al deposition: 20 nm is deposited at 1 nm/s at an angle of $23.83^{\circ}$ 4. 10 minutes of oxidation at 100 mBar 5. Second Al deposition: 40 nm is deposited at 1 nm/s at an angle of $-23.83^{\circ}$ 6. 20 minutes of oxidation at 10 mBar (capping). \textit{Resist Liftoff:} Chip is bathed in acetone for 3 hours at $60^{\circ}$C. Then it is sonicated in the acetone for 5 seconds, followed by 10 seconds of sonication in IPA. Finally, chip is blown dry with $N_2$.

\subsubsection{Carrier Chip}

The process is based on SEEQC’s two-niobium-superconducting-layers recipe for quantum applications \cite{Yohannes2023HighCircuits}. The process involves substrate preparation, niobium (Nb) sputtering, plasma enhanced chemical vapor deposition (PECVD) of $\mathrm{SiO}_2$/$\mathrm{SiN}_x$, chemical mechanical polishing (CMP), photolithography, and dry etching. The substrate is a high resistivity 6 inch (150 mm) silicon wafer. 

It is first cleaned by spin scrubbing with acetone and spin rinse drying with IPA. Then the native oxide on the wafer is removed by BOE prior to loading into the deposition chamber. Once the chamber reaches a base pressure of $1.3 \times 10^{-7}$ mBar, Nb is deposited by DC sputtering with target thickness  200 nm. The coplanar microstrip lines, feedthroughs, resonators, and qubit control lines are then pattered in this layer. All the patterning for this process is done using a maskless laser writer capable of resolving $600~\mathrm{nm}$ features with an i-line resist process. Then the Nb is inductively coupled plasma (ICP) etched in chlorine chemistry optimized for good selectivity to silicon. Moreover, an endpoint detection was implemented to minimize the overetch into the substrate and to keep the overetch to within $10$ nm. Next, the resist is stripped in an acetone sonicating bath, followed by 600 nm of low loss PECVD $\mathrm{SiN}_x$ deposited with CMP to planarize the layer. The wafer is then coated with PECVD $\mathrm{SiO}_2$ of thickness 150 nm for an inter-layer dielectric. The vias to the bottom Nb layer are then patterned and dry etched, which opens access to the wring, grounds, bump and contact pad connections. Then a second Nb layer of thickness 300 nm is deposited, followed by patterning of the wiring, ground straps, and contacts. The Nb is then dry etched in reactive ion etcher (RIE); at this step the exposed dielectric is completely dry etched from the wafer to minimize the dielectric participation ratio. After this step, there is only dielectric remaining under the wiring and ground straps. Next, the contact pads are patterned: trilayer Mo/Pd/Au of thickness 40 nm/100 nm/200 nm is deposited and lifted off.  Finally, the bump layer is patterned: Al/In of thickness 5 $\mu$m/1 $\mu$m is deposited and lifted off. The carrier chips of size 10 $\times$ 10 $\text{mm}^2$ are then diced out and ready for flip-chip bonding.

\subsubsection{Multi-Chip Module}

The multi-chip module was assembled using a Karl Suss / Microtec FC 150 Flip-Chip Bonder. The tool holds the chip and carrier on two vacuum chucks while adjusting the relative position of alignment marks viewed through a two-objective microscope assembly. The bonder was calibrated immediately before MCM assembly, such that the lateral misalignment between chip and carrier is within 1 $\mu$m. The chip and carrier are aligned parallel to within 25\,$\mu$rad, as specified by the tool.

The MCM was assembled by a cold compression process with a force of 294\,$\text{N}/\text{mm}^2$, held constant for 2 minutes. Distributed over the 152 square bumps, each 120\,$\mu$m wide (8.8\% of the total flip-chip area), this force is equivalent to 197 grams per bump. Post-bond measurement of bump height on other MCMs has indicated that this force compresses the top 1\,$\mu$m layer of indium, but does not significantly compress the underlying aluminum bump, such that the aluminum acts as a hard stop to control inter-chip spacing. After bonding, a small amount of low-temperature microelectronic-compatible epoxy was applied to the corners of the chip, away from critical circuit components, to maintain integrity of the MCM during sample handling and installation.

\subsection{Experimental Setup}
\label{app_exp}

\begin{figure}[h]
	\centering
	\includegraphics[width=.92\linewidth]{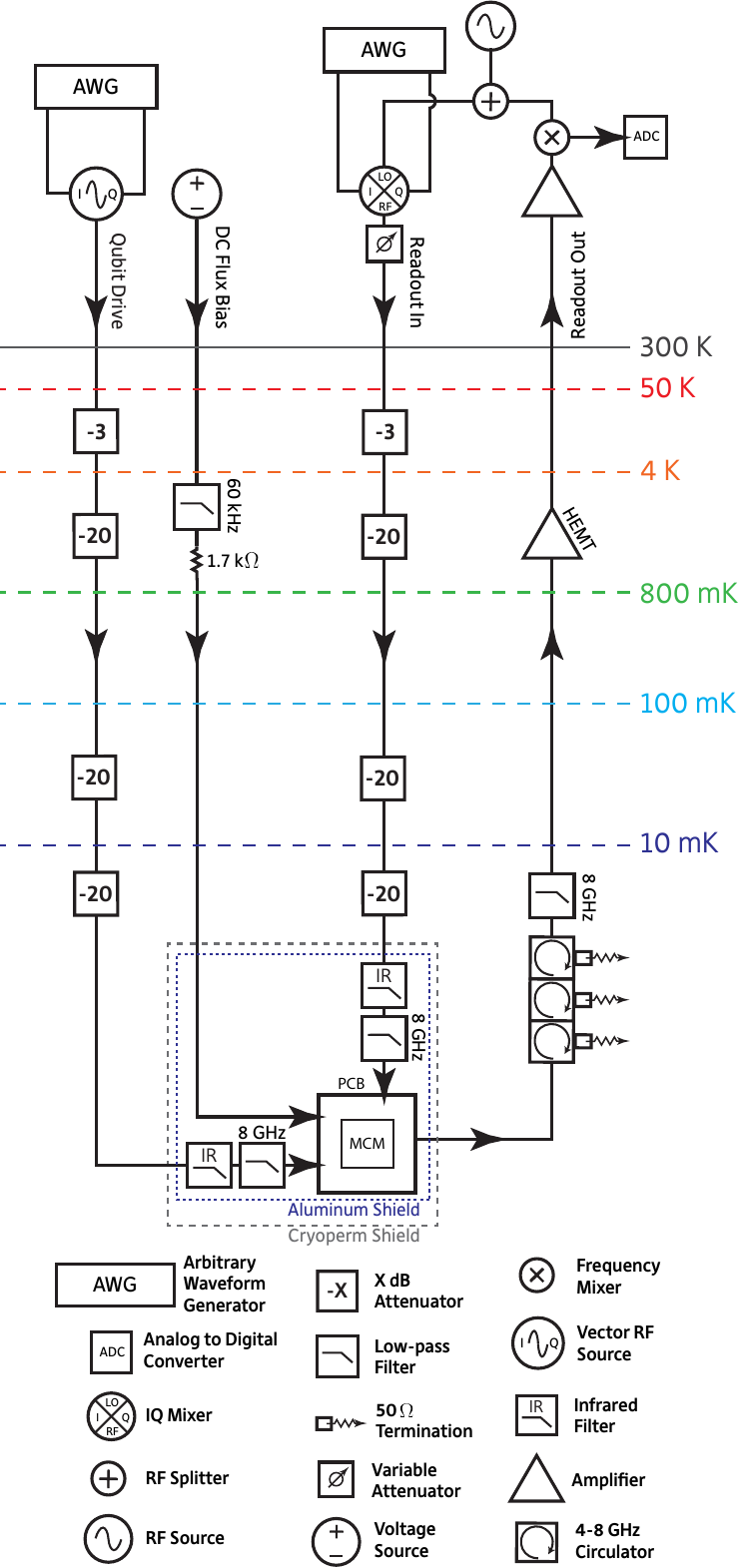}
	\caption{Schematic of the wiring in the DR and room temperature setup. The qubit drive and readout input lines are heavily attenuated and filtered using 63 dB of commercial attenuation (BlueFors cryogenic attenuators), along with K\&L low-pass filters and infrared (IR) filters (manufactured by BlueFors). 
	The DC flux bias line consists of a twisted pair, the positive terminal set by a voltage source and the negative terminal connected to the common ground plane of the chip. Each component in the figure is thermally anchored to the stage directly above it.}
	\label{fig:exp_setup}
\end{figure}	

The experiment was performed in a BlueFors LD400 dilution refrigerator (DR) at a base temperature of $10-20 \mathrm{~mK}$. The driving and readout lines were attenuated and filtered as depicted in Figure \ref{fig:exp_setup}. The MCM was mounted and wirebonded to a printed circuit board (PCB), which in turn was housed in a copper package. This package was bolted to a coldfinger extending from the 10~mK stage of the DR. The device package was encased in a superconducting (aluminum) shield as well as a cryoperm shield. Additionally, we used a magnetic shield inside the outer vacuum can of the refrigerator.

\clearpage

%

\end{document}